\documentstyle[aps]{revtex}

\input epsf

\begin{document}
\thispagestyle{empty}
\baselineskip 18pt

\rightline{KIAS-P02035}
\rightline{SNUTP-02018}
\rightline{ {\tt hep-th}/0205274}

\def\sdtimes{\mathbin{\hbox{\hskip2pt\vrule
height 4.1pt depth -.3pt width .25pt\hskip-2pt$\times$}}}

\def\tr{{\rm tr}\,}
\newcommand{\beq}{\begin{equation}}
\newcommand{\eeq}{\end{equation}}
\newcommand{\beqn}{\begin{eqnarray}}
\newcommand{\eeqn}{\end{eqnarray}}
\newcommand{\bde}{{\bf e}}
\newcommand{\balpha}{{\mbox{\boldmath $\alpha$}}}
\newcommand{\bsalpha}{{\mbox{\small\boldmath$alpha$}}}
\newcommand{\bbeta}{{\mbox{\boldmath $\beta$}}}
\newcommand{\blambda}{{\mbox{\boldmath $\lambda$}}}
\newcommand{\ggg}{{\boldmath \gamma}}
\newcommand{\ddd}{{\boldmath \delta}}
\newcommand{\mmm}{{\boldmath \mu}}
\newcommand{\nnn}{{\boldmath \nu}}

\newcommand{\bra}[1]{\langle {#1}|}
\newcommand{\ket}[1]{|{#1}\rangle}
\newcommand{\sn}{{\rm sn}}
\newcommand{\cn}{{\rm cn}}
\newcommand{\dn}{{\rm dn}}
\newcommand{\diag}{{\rm diag}}

\

\vskip 1cm

\centerline{\Large\bf 1/4 BPS Dyoninc Calorons}

\vskip 1.5cm \centerline{\large\it Kimyeong
Lee$^{a\natural}$ and Sang-Heon
Yi$^{b\dagger}$}

\vskip 5mm

\centerline{$^a$School of Physics, Korean Institute for Advanced
Study} \centerline{207-43, Cheongryangri-Dong, Dongdaemun-Gu,
Seoul 120-012, Korea}
\vskip 5mm
\centerline{$^b$Department of Physics and CTP, Seoul National
University,
 Seoul  151-742, Korea}

\vskip 1.5cm
\centerline{\bf ABSTRACT} \vskip 5mm

\begin{quote} {\baselineskip 16pt We explore the 1/4 BPS  
configurations of the supersymmetric gauge theories on $R^{1+3} \times
S^1$.  The BPS bound for energy and the BPS equations are obtained and
the characteristics of the BPS solutions are studied.  These BPS
configurations describe electrically charged calorons, which are
constituted of dyons and carry linear momentum along the compact
direction.  We carry out various approaches to the single caloron case
in the theory of $SU(2)$ gauge group.}
\end{quote}

\vskip  4cm

~\newline $\overline{\mbox{Electronic Mail:} \; {}^\natural
\mbox{klee@kias.re.kr}, \; {}^\dagger \mbox{shyi@phya.snu.ac.kr}} $

\newpage
\setcounter{footnote}{0}

\section{Introduction}

It has been known for sometime that instantons on $R^3\times S^1$,
or so-called calorons, can be considered to be made of magnetic
monopoles when there is a nontrivial Wilson loop which breaks the
gauge symmetry to its abelian subgroups~\cite{piljin,kraan,lu}.
For the $SU(N)$ gauge group, there are $N$ different kinds of
fundamental monopoles, each corresponding to the roots in the
extended Dynkin diagram. The relation between instantons and
magnetic monopoles can also be understood by exploring the five
dimensional Yang-Mills theories which appears as the low energy
Lagrangian on parallel D4 branes and its T-dual version. In their
$N=4$ supersymmetric version on $R^{1+3} \times S^1$, instantons
appear as 1/2 BPS objects.  The low energy dynamics of calorons or
instantons are given by the metric on the moduli space of caloron
solutions.

There has been some works done some time ago on the 1/4 BPS dyons on
$N=4$ supersymmetric Yang-Mills theories on $R^{1+3}$, which can arise
when several Higgs fields takes expectation
values~\cite{leeyi,sasakura}. In the five dimensional Yang-Mills
theories, there are five Higgs fields and they can take nontrivial
expectation values, besides the nontrivial Wilson loop along the
compact circle. In those theories also the 1/4 BPS and non BPS
configurations are possible. In this paper we explore these 1/4 BPS
configurations. Especially we work out a single 1/4 BPS dyonic caloron
case in the $SU(2)$ gauge theory.

More recently there has been some work on dyonic or electrically
charged instantons in five dimensional field
theory~\cite{tong,townsend,zama}. As in the four dimensional theory,
these dyonic instantons are 1/4 BPS instead of 1/2 BPS in the
Yang-Mills theories with sixteen supersymmetries. These dyonic
instantons also carry nontrivial angular momentum. They would become
caloron when the space is compactified. As we will show in this paper,
BPS calorons come with richer characteristics.

Usually we consider a BPS configuration to be at rest. However, they
remain BPS when the configuration is Lorentz boosted. As our space is
compactified along a circle, a BPS configuration can carry nonzero
linear momentum along the circle. However the linear momentum is not
topological  and cannot be expressed as the boundary term in
general.

The 1/4 BPS dyons can be understood as a planar web of fundamental
strings and D-strings connecting parallel D3 branes in type IIB
theory~\cite{bergman}. Similarly the 1/4 BPS dyonic calorons have
the string web picture. We explore this in our simple model.

The low energy dynamics of magnetic monopoles can be approached by
the moduli space dynamics. When an additional scalar field is
turned on, its effect can be incorporated as a potential term. It
was shown in~\cite{clee} that the BPS configuration of this low
energy dynamics corresponds to the 1/4 BPS dyonic configurations.
From this correspondence one can read the electric charge of dyons
for a given set of moduli parameters. This result can be also
found directly from the field theory analysis. We consider the low
energy dynamics of 1/4 BPS dyonic caloron  and work out in detail
these results in the $SU(2)$ case.

The plan of this paper is as follows. In Sec.2, we find the BPS bound
on the energy functional. In Sec.3, we find the BPS equations which
are satisfied by the configurations saturating the BPS bound. In
Sec.4, we find the BPS caloron configurations, which can be regarded
to be composed of  monopoles and dyons. In Sec.5, we study the $SU(2)$
gauge group case in detail. Especially, we relate our 1/4 BPS
configuration with the string web picture. In Sec.6, we conclude with
some remarks.

\section{The BPS Bound}

The underlying  spacetime is chosen to be five dimensional, with
one of the space being compactified to a circle. The coordinates,
$x^M$, where  $M=0,...,4$, are split to the time coordinate $x^0$
and space coordinates $x^\mu$ with  $\mu=1,2,3,4$. The
compactified coordinate $x^4$ has the finite range
\beq
0\le x^4< \beta.
\eeq
We consider only the periodic gauge and scalar fields.  The allowed
gauge transforms are those which leave the gauge fields periodic:
small gauge transformations whose gauge functions  are periodic, and large
gauge transformations whose gauge functions are multi-valued.   While
our consideration can be easily generalized to arbitrary semi-simple gauge
group, we focus on the $SU(N)$ gauge group for simplicity.  We
consider the Hermitian generators $T^a$ in $N$ dimensional fundamental
representation with normalization $\tr T^a T^b = \delta^{ab}/2$.  The
gauge field is then $A_M = A_M^a T^a.$

The Lagrangian  we start with is
\beq
{\cal L} = \frac{1}{e^2}\tr \left(-\frac{1}{2} F_{MN}F^{MN} +
D_M\phi_I D^M \phi_I - \sum_{I<J} \bigl(-i[\phi_I,\phi_J]\bigr)^2
\right),
\eeq
where $D_M \phi_I = \partial_M\phi_i - i\big[A_M,\phi_I\big]$ and
$e^2$ is the five dimensional coupling constant of length
dimension.
%
%We will choose the dimensionful five dimensional coupling constant
%$e=1$ for the convenience.
%
We decompose the Higgs field into one component and the rest,
\beq
\phi_I = a_I \phi + \zeta_I,
\eeq
where $a_I$ is a unit vector in five dimension and $\zeta_I$ is
orthogonal to $a_I$.
The Gauss law is
\beq
D_i E_i + D_4 F_{04} -i[\phi,D_0\phi]-i[\zeta_I, D_0\zeta_I]=0,
\label{gauss1}
\eeq
where $E_i = F_{0i}$ with $i=1,2,3$.

The energy density is given by
\beq {\cal E} = \frac{1}{e^2}\tr \left( E_i^2+ B_i^2 + F_{04}^2+
F_{i4}^2 + D_0\phi^2 +D_i\phi^2 + D_4\phi^2 \right) + {\cal
E}_\zeta , \eeq
where $B_i = \frac{1}{2}\epsilon_{ijk}F_{jk}$ and
\beq
e^2 {\cal E}_\zeta = \tr \left( (D_0\zeta_I)^2 + (D_\mu \zeta_I)^2 -\sum_I
[\phi,\zeta_I]^2 -\sum_{I<J}[\zeta_I,\zeta_J]^2
\right) . \eeq
The energy density can be written as
\beqn {\cal E} &=& \frac{1}{e^2}\tr \left\{ (E_i + F_{4i}  \sin
\alpha - D_i \phi \cos \alpha )^2
+ (B_i - F_{4i} \cos \alpha - D_i \phi \sin \alpha )^2 \right.\nonumber\\
& & \;\;\;  \;\;\; \left.
+ (F_{04}-  D_4 \phi\cos \alpha )^2
+ (D_0\phi + D_4 \phi\sin \alpha )^2 \right\} \nonumber \\
& & + 2\cos \alpha\; \left( \tr   B_i F_{4i} + \partial_i \tr (E_i
\phi) \right) \nonumber \\
 & & + 2\sin\alpha \;   \left\{ \partial_i\tr (B_i \phi) - \tr (
E_iF_{4i}+D_0\phi D_4\phi + D_0\zeta_I D_4 \zeta_I ) \right\} +
\tilde{\cal E}_\zeta ,
\eeqn
where
\beqn
e^2 \tilde{\cal E}_\zeta = &&  \tr \left( (D_0 \zeta_I -i [\phi,
\zeta_I] \cos \alpha  +D_4 \zeta_I  \sin \alpha  \right)^2 +
\tr \left( D_4 \zeta_I \cos \alpha  + i [\phi, \zeta_I] \sin\alpha
\right)^2 \nonumber \\
&& + (D_i \zeta_I)^2 -
\sum_{I<J} [\zeta_I,\zeta_J]^2.
\eeqn
In the above we have used the Gauss law and the single-valuedness
of the fields in $x_4$.

We introduce  four conserved charges;
\beqn
&& Q^E = \frac{2}{e^2}\int d^4x  \; \partial_i \tr(E_i\phi) \; , \\
&& Q^M = \frac{2}{e^2}\int d^4x \; \partial_i \tr(B_i\phi) \; , \label{qm} \\
&& P^4 = -\frac{2}{e^2}\int d^4x\; \tr (E_i F_{4i} + D_0\phi D_4\phi +
D_0 \zeta_I
D_4 \zeta_I) \label{p4}             \; ,\\
&& {\cal T} =  \frac{8\pi^2}{e^2} \nu_P \; ,
\eeqn
where $d^4x $ is the volume element of the four dimensional space.
The linear momentum along the circle $P^4$ is conserved but is
not topological. The rest of them are topological. Especially
${\cal T}$ is the related to the Pontriyagin index by
\beq
\nu_P = \frac{1}{8\pi^2} \int d^4x \, 2\,\tr (B_i
F_{4i}) \; .
\eeq
A  bound on the energy functional $H= \int d^4 x {\cal E}$ is then
\beq
H \ge ({\cal T} + Q^E) \cos \alpha  + (Q^M + P^4)  \sin\alpha
\eeq
for any angle $\alpha$. The maximum possible value of the right side
is obtained when
\beqn
&& \cos \alpha = ({\cal T} + Q^E)/\sqrt{({\cal T}+Q^E)^2 +
(Q^M+P^4)^2} \; , \\
&& \sin \alpha = (Q^M+P_4)/\sqrt{({\cal T}+Q^E)^2 + (Q^M+P^4)^2}
\;  \label{sinalpha}.
\eeqn
Then the strictest energy bound is then
\beq H \ge Z_\pm= \sqrt{({\cal T}\pm Q^E)^2 + (Q^M\pm P^4)^2} \; ,
         \label{bpsbound}
\eeq
which is the so-called BPS energy bound. $Z_-$ is obtained by changing
the sign of $\phi$ field.

%--------------------------------------------------------------------------
\section{BPS Equations}
%--------------------------------------------------------------------------

This energy bound is saturated, say, $H = Z_+$, by the configurations
which satisfy the  following BPS equations,
\beqn
&& B_i =  F_{4i} \cos\alpha +  D_i\phi\sin\alpha , \label{bps1} \\
&& E_i = -F_{4i} \sin\alpha  + D_i\phi \cos\alpha ,\label{bps2} \\
&& F_{04}  = D_4 \phi \cos\alpha  ,  \label{bps3} \\
&& D_0\phi = - D_4 \phi \sin\alpha , \label{bps4}
\eeqn
and the Gauss law (\ref{gauss1}).  In addition, there are conditions
for the $\zeta_I$ field,
\beqn
&&D_0 \zeta_I - i [\phi,\zeta_I] \cos \alpha +   D_4 \zeta_I \sin \alpha
= 0 , \label{bpsz1}\\
&& D_4\zeta_I  \cos\alpha + i [\phi,\zeta_I] \sin\alpha = 0 , \label{bpsz2}\\
&& D_i \zeta_I = 0 , \label{bpsz3}\\
&& [\zeta_I, \zeta_J] = 0 . \label{bpsz4}
\eeqn
The above BPS equations seem to be complicated to solve. However,
a considerable simplification can be made by noticing
Eq.~(\ref{bps1}) can be written as a self-dual equation. Let us
introduce a new coordinate
\beq
\tilde{x}^4 = \frac{1}{\cos\alpha}x^4 ,
\eeq
and a new fourth component gauge field
\beq
\tilde{A}_4 =A_4   \cos\alpha  - \phi \sin\alpha .
\eeq
Then the equation (\ref{bps1}) becomes
\beq
B_i = \tilde{\partial}_4 A_i - D_i \tilde{A}_4 = \tilde{F}_{4i} .
\label{bpse1}
\eeq
This is the self-dual equation for calorons on $R^3 \times
\tilde{S}^1$, where $\tilde{x}^4 \in [0,\beta/\cos\alpha]$.  Since the
fields are periodic under $x^4$, they are period under $\tilde{x}^4.$
If $\alpha=\pi/2$, the above method fails. However, the BPS equations
in this case become those of 1/2 BPS monopoles with a single scalar
field $\phi$ and studied extensively. (Not all 1/4 BPS configurations
of the theory is not described by the above BPS equation. For those
configurations, two Higgs fields  are  involved and $A_4=0$,  and so
the compactified direction does not play any role. )

Introducing
\beq
\tilde{D}_4 = \tilde{\partial}_4- i \tilde{A}_4,
\eeq
Eq.~(\ref{bps3}) becomes $F_{04} = \tilde{D}_4 \phi$ and
Eq.~(\ref{bps4}) becomes $D_0\phi = - \tilde{D}_4 \phi \tan
\alpha $. Taking a covariant divergence of Eq.~(\ref{bps1}), we get
\beq
D_i F_{4i} \cos\alpha  + D_i^2\phi \sin\alpha = 0
\eeq
Using the above relations with Eqs.~(\ref{bpsz1}), (\ref{bpsz2}),
Eq.~(\ref{bps2}) and the Gauss law (\ref{gauss1}) can be put in a
single equation,
\beq
D_i^2 \phi + \tilde{D}_4^2 \phi - [\zeta_I,[\zeta_I,\phi]]=0.
\label{bpse2}
\eeq
Eq.~(\ref{bpsz2}) can be put in a form $\tilde{D}_4\zeta_I=0$.
Since we are interested in the 1/4 BPS configuration such that the
vacuum expectation values of $\zeta_I$ vanish $(\langle \zeta_I
\rangle=0)$ and $D_i\zeta_I=\tilde{D}_4\zeta_I=0$, $\zeta_I$
should vanish. Hence, we can drop the $\zeta_I$ fields from the further
discussion.
%%%%%%%%%%%%%%%%%%%%%%%%%%%%%%%%%%%%%%%%%%%%%%%%%%%%%%%%%%%%%%%%%%
Thus, Eq.~(\ref{bpse2}) becomes
\beq
D_i^2 \phi + \tilde{D}_4^2\phi= 0 .
\label{bpse22}
\eeq

The topological charge of the new variables
\beq \tilde{\cal T} =\frac{2}{e^2}  \int d^4 \tilde{x}  \tr B_i
\tilde{F}_{4i} = {\cal T} -  Q^M  \tan \alpha \; .
\label{ttrans}
\eeq
Note that $\tilde{\cal T}$ or  ${\cal T}$ need not be integer
number for the generic configuration.  Another topological
quantity appearing naturally here is
\beq
\tilde{Q}^E= \frac{2}{e^2}\int d^4\tilde{x}  \tilde{\partial}_\mu  \tr
\phi \tilde{D}_\mu \phi
= Q^E +  Q^M \tan \alpha
\label{qetrans}
\eeq
for BPS configurations.

\subsection{The Case where $\alpha = 0 $}

What is the reason behind this simplification of BPS equation? Let
us first consider the $\alpha =0$ case. The BPS equations comes
quite simplified:
\beqn
&& B_i = F_{4i} , \label{bpso1} \\
&& E_i = D_i  \phi ,\label{bpso2} \\
&& F_{04} = D_4 \phi , \label{bpso3} \\
&& D_0 \phi = 0 .
\label{bpso4}
\eeqn
In the gauge $A_0 = -\phi$, the field configurations become time
independent. Especially Eqs.~(\ref{bpso3}) and (\ref{bpso4}) are
automatically satisfied.  The fields $A_\mu$ satisfy the self-dual
equation and the Gauss law constraint can be put in a simple form
\beq
D_\mu^2 \phi =  0.
\label{bpso5}
\eeq
{}From the above simplified BPS equations, we can see $P^4 = -Q_M$
from Eqs. (\ref{p4},\ref{qm}), which is consistent with $\alpha=0$
picture in BPS bound (\ref{bpsbound}).

\subsection{Lorentz boost along $x^4$ }

Let us start with the $\alpha=0 $ case.  We call its spacetime
coordinates $\tilde{x}_M$ and its BPS field configurations
$\tilde{A}_M$ and $\tilde{\phi}$.  We can Lorentz boost  this
coordinate to get a new coordinate $(x^4,t)$ such that
\beqn
&& \tilde{t} = \frac{t}{\cos\alpha} -  x^4 \tan \alpha \\
&& \tilde{x}^4 =  - t \tan \alpha +  \frac{x^4}{\cos\alpha} .
\eeqn
Note that $1/(\cos\alpha)^2 - (\tan\alpha)^2= 1$ so that it is a
Lorentz boost.  The spatial coordinates $x^i$ remain unchanged.  We
start with the compact radius $0<\tilde{x}^4<\beta/\cos\alpha$, and so
the compact radius of $x^4$ becomes $\beta$.

The $A_0$ and $A_4$ fields transform as
\beqn
&& \tilde{A}_0 = \frac{A_0 }{\cos\alpha}+  A_4 \tan \alpha ,\\
&& \tilde{A}_4 =  A_0 \tan \alpha + \frac{A_4}{\cos\alpha} .
\eeqn
and the $\phi$ and $A_i$ fields are invariant. We work in the gauge
$\tilde{A}_0=\tilde{\phi}=\phi$. The BPS equations
(\ref{bpso1}$\sim$\ref{bpso4}) of  $\alpha=0$ case becomes the old
BPS equations (\ref{bps1}$\sim$ \ref{bps4}) in terms of new variables.

What is interesting about this Lorentz transformation is that
${\cal T}+ Q^E$ is like the rest mass and $P^4+Q^M$ is like the
four momentum of the Lorentz boost. (The Lorentz boost
interpretation of Eqs.~(\ref{ttrans}) and (\ref{qetrans}) shows
that while ${\cal T}$ and $Q^E$ transform nontrivially ${\cal
T}+Q^E$ remains invariant as the rest mass.)  Thus when
$\alpha=0$, the momentum $P^4+ Q^M =0$ as noted before. Of course
our Lorentz boost is not an exact symmetry as the interval of
space changes and is a kind of a combination of Lorentz boost and
rescaling of the $x^4$ coordinate. However, this shows that the
BPS configuration with nonzero $\alpha$ can be obtained from the
BPS configurations with $\alpha=0$. This is exactly what we have
seen in Eq.~(\ref{bpse1}) and (\ref{bpse22}). While the BPS
configuration of the $\alpha=0$ case can be chosen to be time
independent, this is not true for those of the $\alpha \ne 0$.

\section{ Monopoles and Calorons}

The solutions of the primary BPS equation (\ref{bpse1}) is identical
to the self-dual equations for the caloron. Since we are interested in
the case where $\alpha \ne \pi/2$ and so it can be obtained from the
Lorentz boost from the case $\alpha=0$, we focus mainly on the
$\alpha=0$ case.  The first one (\ref{bpso1}) is the self-dual
equations of $A_\mu$ on $R^3\times S^1$. The second BPS equation is
the zero eigenvalue equation (\ref{bpso5}) for the $\phi$, around the
solution of the first BPS equation.

Let us first consider the solution of the primary  BPS equation. This
is the BPS equation for the 1/2 BPS configurations. First
of all we need the boundary condition for $A_4$.  The vacuum
expectation value of $A_4$ is single-valued and takes the form
\beq
<A_4> = \diag (h_1,h_2,...,h_N)= {\bf h}\cdot{\bf H}\; ,
\eeq
where $\sum_a h_a = 0$ and by gauge choice
\beq
h_1<h_2<...<h_N<h_1+ \frac{2\pi}{\beta}
\label{gaugec}
\eeq
This leads to a nontrivial Wilson-loop $P \exp (i\int dx^4
A_4)$ and the symmetry is spontaneously broken to
$U(1)^{N-1}$. If any of two $h_a$'s are coincide, the gauge
symmetry will have unbroken nonabelian symmetry. While this
possibility is quite interesting by itself, we will not pursue
this direction in the present paper.

As shown in Ref.~\cite{piljin}, the general solutions of the
self-dual equation (\ref{bpso1}) describe the superpositions of N
fundamental monopoles corresponding to the simple roots
$\bbeta_i$, $i=1,...,N-1$ and the lowest negative root $\bbeta_0$.
These roots form the extended Dynkin diagram.  The topological
charge $\nu_P$ of each type of monopole is fractional and takes
the fractional value,
\beqn
&& \mu_r = \frac{\beta}{2\pi} (h_{r+1}-h_r),\;\;\; r=1,...,N-1
\nonumber \\
&& \mu_0 = 1- \frac{\beta}{2\pi} (h_N - h_1)
\eeqn
While all monopoles are on the equal footing, the magnetic monopoles
of $\bbeta_0$ are called Kaluza-Klein monopoles as it has intrinsic
$x^4$ dependence in the gauge (\ref{gaugec}).

Thus the general solution of the primary  BPS equation is characterized
by the $N$ nonnegative integers $n_r$, each of which is the number of
$\bbeta_r$ monopoles.  The total topological charge is then
\beq
\nu_P = \sum_0^{N-1} n_r \mu_r,
\label{nup2}
\eeq
and the magnetic charge obtained from asymptotic of $B_i =
(r_i/r^3){\bf g}\cdot {\bf H}$  is given by
\beq
{\bf g} = 4\pi  \sum_0^{N-1} n_r \bbeta_r.
\eeq
The topological charge and the magnetic charge are $N$ charges
together and so  determine the monopole numbers  $n_r$ $r=0,...,N$
uniquely. The total number of the partons or monopoles is
\beq
N_m = \sum_0^{N-1} n_r,
\eeq
and the total number of zero modes of the first BPS equation is
$4N_m$.  A single caloron or instanton can be regarded as
composite of $N$ distinct fundamental monopoles, whose topological
charge is one and magnetic charge is equal to zero.  The number of
zero mode of a single instanton  is then $4N$, as expected from
the index theorem.  For a given set $\{n_r\}$ of monopoles, the
solution of the first BPS equation is uniquely determined by the
moduli parameters. The dimension of the moduli space of these
configurations is 
$4N_m$.  The general method to solve the first BPS equation is the
ADHMN construction, as detailed in Ref.~\cite{lu}. Thus, the
solutions of the primary BPS equations are identical to 1/2 BPS
configurations.

The secondary BPS equation can be regarded as the gauge zero modes of the
primary  BPS equation.  The linear fluctuations $\delta A_\mu$ should
satisfy the linearized BPS equation and the gauge fixing condition,
\beqn
&& \epsilon_{ijk}D_j \delta A_k = D_4 \delta A_i - D_i \delta  A_4\;,
\nonumber \\
&& D_\mu \delta A_\mu = 0.
\label{zeroe}
\eeqn
When the linear fluctuation is due to the gauge zero mode, $\delta
A_\mu = D_\mu \Lambda$, the linearized BPS equation is automatically
satisfied and the gauge fixing condition becomes
\beq
D_\mu^2 \Lambda = 0,
\label{gpara}
\eeq
which is identical to the secondary BPS equation (\ref{bpso5}). Since
there is $N-1$ unbroken U(1) symmetries, there will be $N-1$ linearly
independent solutions of the second BPS equation. The general method
for solving the second BPS equation is given in the appendix of
Ref.~\cite{leeyi}.

The BPS equation (\ref{bpso5}) gives the electric field in terms of
the solution $\phi$.  The solution of the second BPS equation is again
uniquely determined for a given monopole background and the asymptotic
value,
\beq \langle\phi\rangle= (a_1,a_2,...,a_N)={\bf a}\cdot{\bf H}\;
,\qquad \sum_{i=1}^N a_i=0 \; . \label{asymta} \eeq
Thus, the moduli of the 1/2 BPS configurations determine the solution
of the secondary BPS equation uniquely.
{}From the asymptotics of the field $\phi$
\beq
\phi = \langle\phi\rangle + \frac{1}{4\pi r}{\bf q}\cdot{\bf H} +
{\cal O}(\frac{1}{r^2}) \; ,
\label{asymtb}
\eeq
we can read the electric field and so the electric charge $Q^E$.

However, the story is more complicated.  There are $N$ distinct
monopoles, each of which can carry its own electric fields. Thus for a
group of $\bbeta_r$ monopoles, there will be total $q_r \bbeta_r$
electric charge and the electric charge would be determined by the
asymptotics of $\phi$ field is given by
\beq
{\bf q} = \sum_0^{N-1} q_r \bbeta_r \; .
\label{elecf}
\eeq
Similarly to the magnetic charge, the asymptotics alone cannot decide
each $q_r$ since $\bbeta_r$ is not independent although it determines
the relative electric charges. The electric charge ${\bf q}$ is
determined by the moduli parameters of the magnetic background and the
asymptotic value ${\bf a}$.

Besides the $N-1$ global $U(1)$ abelian symmetries of the gauge group
$SU(N)$, there is an additional $U(1)$ related the translation along
$x^4$ direction. Thus there are $N$ global $U(1)$ charges which
consist of the $N-1$ electric charge and the linear momentum $P^4$,
which in turn determines the $N$ parameters $q_r$. The linear momentum
$P^4$ is not topological and so it is hard to see its relation to
$q_r$ explicitly. When constituent monopoles are well-separated from
each other, one may be able to assign electric charge and linear
momentum to each monopole.  Especially when $\alpha=0$, the linear
momentum (\ref{p4}) for very isolated $\beta_r$ monopoles of electric
charge $q_r$ carries linear momentum $q_r\mu_r$.  Then the total
linear momentum becomes $\sum_r q_r \mu_r$, which should be $-Q^M$ due
to the equation (\ref{sinalpha}).  This leads an additional relation
between $q_r$'s. The configuration with nonzero $\alpha$ can be
obtained from the Lorentz boost of the above case.

%----------------------------------------------------------------------
\section{ Low Energy Dynamics of Calorons}
%----------------------------------------------------------------------

{}From the consideration in the previous sections, one can see that 1/4
BPS dyonic calorons can be constructed of fundamental dyons. The low
energy dynamics of 1/4 BPS configurations has been explored in recent
years~\cite{clee,bak1}. It was shown that the low energy dynamics is
possible for  1/2 BPS configurations. The kinetic energy is given by
the moduli space metric of the 1/2 BPS configurations.  The additional
Higgs field appearing in the 1/4 BPS configuration contributes to the
potential to the low energy Lagrangian. The form of the potential is
given by the sum of the square norm of Killing vectors. For the low
energy dynamics of fundamental objects one has to specify what class
of 1/2 BPS configurations one starts with. For example one can start
from the case where $\langle A_4 \rangle \ne 0$ and $\langle
\phi\rangle = 0$. In this case one starts with the constituent
fundamental monopoles of calorons. Then we consider the low energy
dynamics of monopoles with small but nonzero $\phi$
expectation value. This is the case we will focus in this work.

There are other cases which we will not explore here. With
 $\langle A_4 \rangle = 0$ and $\langle \phi \rangle \ne 0$, one start
 with the the 1/2 BPS monopoles  without any
Kaluza-Klein monopoles, which would make calorons.  It would be
interesting to consider the low energy dynamics of these monopoles
with small but nonzero $\langle A_4 \rangle$.  One could consider
more complicated 1/2 BPS configurations with $\langle A_4 \rangle
\propto\langle \phi \rangle$ but without Kaluza-Klein monopoles.

The moduli space dynamics of dyonic instantons on $R^4$ has been
studied before in Refs.~\cite{tong,peet}. It is somewhat  simpler than
our case as there is no symmetry breaking due to the $A_4$ expectation
value.

\subsection{Caloron Moduli Space}

We start with the caloron case with $\langle \phi\rangle =0$. Its
low energy dynamics would be described by the caloron moduli space
metric. The kinetic energy due to the spatial motion and electric
charge would be much smaller than the rest mass of monopoles.  We
consider the modification of the low energy dynamics when $\langle
\phi \rangle \ne 0$ but very small. Thus the angle $\alpha$ would
be very small. The $\alpha\neq 0$ case can be obtained by the
Lorentz transformation and rescaling of the $x^4$ coordinate. As
$\alpha$ is very small, it would become an infinitesimal Galilean
transformation along $x^4$ direction. Under this transformation
the  electric field transforms more  complicatedly as one can see
from Eq.~(\ref{bps2}). The moduli space metric gets split to that
for the center of mass motion and that for the relative motion.
Thus we just focus on the case $\alpha=0$.

In this case the solution of the primary BPS equation is identical to
the 1/2 BPS caloron solutions.  We assume that all $n_r$ are positive
and so all constituent monopoles are interacting. In this case, the
1/2 BPS configuration has intrinsic $x^4$ dependence which cannot be
gauged away.

The dimension of the moduli space is $4N_m$ and the moduli space
coordinates are $z^M$ with $M=1,...,4N_m$. The moduli space metric
can be obtained from the study of the linear fluctuations around
the 1/2 BPS configurations $A_\mu(x,z)$,
\beq
\delta_M A_\mu = \frac{\partial A_\mu }{\partial z^M} - D_\mu \epsilon_M
\eeq
which satisfies Eq.~(\ref{zeroe}). The moduli space metric is
\beq
g_{MN}(z) = 2 \int d^4x\, \tr (\delta_M A_\mu \delta_N A_\mu )
. \label{metric}
\eeq

For  instanton or caloron solutions of $SU(N)$ gauge
theory, the moduli space is $4N$ dimensional and has orbifold
singularity at the point where all monopoles come together and the
caloron collapses. This moduli space metric was obtained using the
constituent monopole picture and the nature of singularity is
identified as $R^{4N}/ Z_N$~\cite{piljin}.  It was also known that
the singularity is resolved if we turn on the
non-commutativity~\cite{leeyi2}.

The effective Lagrangian for the low energy dynamics can be
generically written as
\beq
L_K = \frac{1}{2e^2} g_{MN} \dot{z}^M \dot{z}^N \; .
\label{kenergy}
\eeq
There is a natural $N=4$ supersymmetric generalization of the above
Lagrangian.  Since there are $N$ fundamental monopoles in calorons,
instead of $N-1$ for $SU(N)$ gauge theory, there are $N$ conserved
$U(1)$ symmetries, each leading to the $q_r \bbeta_r$ electric charges
on $\bbeta_r$ monopoles. This matches with the field theoretic
symmetires, $N-1$ of them are made of unbroken abelian subgroups of
$SU(N)$ and one is for the translation along $x^4$ direction.

To understand the dynamics better, let us split the center of mass
motion and the relative motion. The Lagrangian (\ref{kenergy}) would
become the sum of the Lagrangian $L_{K\; cm}$ for the center of mass
motion and $L_{K\; rel}$ for the relative motion.

As the position of individual monopole is not well defined when two
identical monopoles are coming together, it is hard to express the
center of mass position in general. However one can argue that
the  charge for the central $U(1)$ should be
\beq
q_{cm} = \frac{1}{\nu_P} \sum_{r=0}^n q_r \mu_r
\eeq
It is true in large separation. (See for example Ref.~\cite{klee2}.)
Also $q_{cm}$ should be a linear combination of $q_r$ and is
conserved, and so the coefficient should be independent of the
monopole positions. We would like to identify $P_4
=\frac{2\pi}{e^2}\nu_P q_{cm} $ in large separation.  As we consider
the $\alpha=0$ case, the center of mass charge would be constrained to
be $-Q_M =- \frac{1}{e^2}{\bf a}\cdot {\bf g}$.  The nonzero $\alpha$
case is obtained by the infinitesimal Galilean transformation along
the $x^4$ direction. Eq.~(\ref{sinalpha}) implies $P_4 + Q_M =
\frac{8\pi^2}{e^2} \nu_P \alpha$ with $\nu_P$ in Eq.~(\ref{nup2}).
The Killing vector on the moduli space corresponding to the $x^4$
translation would be $K_{cm} = \partial/\partial \psi_{cm}$ with the
center of mass angle variable $\psi_{cm}$.

The relative charges arise as the Noether charges for the $N-1$
Killing vectors $K_r^M \partial_M$ with $r=1,2,...,N-1$ on the moduli
space, which correspond to the $N-1$ unbroken generators of
$SU(N)$. For each of this Killing vector, there would be a cyclic
coordinate $\psi_r$.

We can separate the moduli space coordinates $z^M$ to $({\bf
r}_{cm}, \psi_{cm})$ for the center of mass motion and
$(y^i,\psi^r)$ with $r=1,2,..,N-1$ for the relative motion. The
kinetic energy (\ref{kenergy})  also get split to  $L_{cm}({\bf
r}_{cm}, \psi_{cm})$ for the center of mass motion and
$L_{rel}(y^i,\psi^r) $ for the relative motion. The center of mass
motion is free and trivial.  The relative part $L_{rel}$ is
independent of cyclic coordinates $\psi_r$  and can be expressed
as
\beq
L_{rel} = \frac{1}{2e^2}h_{ij}(y) \dot{y}^i \dot{y}^j + \frac{1}{2e^2}
L_{rs}(y)( \dot{\psi}^r + w^r_i(y) \dot{y}^i) ( \dot{\psi}^s + w^s_j(y)
\dot{y}^j) \; . \label{llag}
\eeq
The $N-1$ conserved abelian charges of the above  Lagrangian   is
\beq t_r = \frac{1}{e^2} L_{rs}(y) (\dot{\psi}^s+ w^s_i(y)\dot{y}^i )
\eeq

%--------------------------------------------------------
\subsection{The Potential}
%--------------------------------------------------------

For a given 1/2 BPS caloron background specified by its moduli
parameters, we introduce the scalar field $\phi$ which takes nonzero
but very small expectation value. The scalar field would take the
lowest possible energy when it satisfy $D_\mu^2 \phi=0$. This scalar
field in turn modifies the caloron dynamics in the second order in
$\phi$. Especially its asymptotic behaviour would be given by
Eqs.~(\ref{asymta}) and (\ref{asymtb}). Especially ${\bf q}=\sum_r
(q_r-q_0) \bbeta_r = \sum_{r=1} q_r^{rel}\bbeta_r $ would depend on
${\bf a}$ and the caloron moduli parameters $z^M$. At this moment
${\bf q}\cdot {\bf H}$ are not electric charge but simply the $1/r$
piece of the asymptotic $\phi$.

The solution of $D_\mu^2\phi=0$ would depend on the caloron moduli
parameters. As $\phi$ satisfies the same equation (\ref{gpara}) as the
global gauge transformation parameters, we can interpret
\beq
D_\mu \phi = {\bf a}\cdot {\bf K}^M \delta_M A_\mu
\eeq
where ${\bf K}^m\partial_M = \sum_{r=1}^{N-1} \bbeta_r K_r^M
\partial_M$ are $N-1$ Killing vectors for the $N-1$ $U(1)$
gauge generators.   As the global gauge transformations change the
cyclic coordinates $\psi^r$ of the relative motion, we choose these
cyclic coordinates such that
\beq
K_r^M \frac{\partial}{\partial z^M } = \frac{e^2}{\beta}
\frac{\partial}{\partial \psi^r}
\eeq
Then the part of the relative moduli space metric becomes $L_{rs}(y) =
(e^4/\beta^2) g_{MN}K^M_r K^N_s$.  The induced potential
energy~\cite{Tong1,clee} is then given by
\beqn
{\cal U} &=& \frac{1}{e^2} \int d^4x \tr (D_\mu\phi)^2 \nonumber \\
&=& \frac{\beta}{2e^2}{\bf q}\cdot {\bf a} \nonumber \\
&=& \frac{1}{2e^2} g_{MN}
{\bf a}\cdot {\bf K}^M {\bf a}\cdot {\bf K}^N \nonumber \\
&=&  \frac{\beta^2}{2e^6} L_{rs} a^r a^s
\eeqn
where ${\bf a} = \sum_r a^r \blambda_r$ with the fundamental weights
satisfying $\blambda_r \cdot \bbeta_s = \delta_{rs}$. The Tong
formula~\cite{Tong1} fixes the asymptotic value ${\bf q}$ explicitly
in terms of the moduli space metric and the asymptotic value
\beqn
\beta {\bf q} &=& g_{MN} {\bf a}\cdot {\bf K}^M {\bf  K}^N  \\
&=& \frac{\beta^2}{e^4} L_{rs} a^r \bbeta_s
\eeqn

The Hamiltonian, $H_{rel}= L_{rel}  +
{\cal U} $ has a BPS bound because 
\beq
H = \frac{1}{2e^2} h_{ij}(y)\dot{y}^i \dot{y}^j
+  \frac{e^2}{2} L^{rs}(y)(t_r \mp \frac{\beta}{e^4} L_{rr'}a^{r'})(t_s
\mp \frac{\beta}{e^4}L_{ss'}a^{s'})
\pm \frac{\beta}{e^2} q_r a^r \; ,
\eeq
which is saturated when $\dot{y}^i=0$ and
\beq
t_r =  q^{rel}_r \equiv  \frac{\beta}{e^4} L_{rs} a^s
\eeq
The energy for the BPS configuration is then
\beq
\frac{\beta}{e^2} a^r q^{rel}_r  = \frac{\beta^2}{e^6}  L_{rs}(y) a^r a^s \; .
\eeq
The above results for the electric charge and energy for the BPS
configuration match exactly that from the field theoretic one
(\ref{elecf}) with $\alpha=0$, and also the above  BPS energy plus the
rest mass ${\cal T}$ becomes the field theoretic BPS energy
(\ref{bpsbound}).

%-------------------------------------------------------------------------
\section{The SU(2) Case}
%-------------------------------------------------------------------------

We explore more in detail the SU(2) case. This shows the above
description of 1/4 BPS configurations in a more concrete form.
The first BPS equation describes the self-dual calorons, and can be
solved by the ADHMN method.  The calorons in the SU(2) case has been
studied in detail before~\cite{lu,kraan}.  Using the small and large gauge
transformations, we choose the expectation value
\beq
<A_4> = -\frac{v}{2}\sigma^3=\diag (h_1,h_2)
\eeq
with $ 0<v\le \pi/\beta$. There are two fundamental monopoles of
opposite magnetic charge. One is ordinary BPS monopole of
topological charge $\mu_1 = \beta v/2\pi$. Another is the KK
monopole of opposite magnetic charge and topological index $\mu_0
= 1-\mu_1$. There is no force between these distinct monopoles as
the magnetic attraction is cancelled by the Higgs repulsion. A
single caloron of a unit Pontriyagin index is made of one of these
monopoles, so that the net magnetic charge is equal to zero. A
single caloron configuration has been obtained explicitly. This
complicated configuration describes nonlinear superposition of two
distinct monopoles.

Consider the $\alpha =0 $ case. The solution of the first BPS equation
is characterized by the two nonnegative integers $(n_0,n_1)$. The
lowest root is $\bbeta_0=-\bbeta_1$ and so the magnetic charge is
\beq
{\bf g} =4\pi (-n_0+n_1)\bbeta_1
\eeq
and the total topological charge is
\beq
\nu_R = n_0 \mu_0 + n_1 \mu_1
\eeq
For a single caloron $n_0=n_1=1$, the solution of the primary BPS
equation has been found by the ADHMN method and was shown to be
superposition of two these monopoles.  Furthermore, the moduli space
of the 1/2 BPS configurations was shown to be 8 dimensional such that
the center of mass part is flat space and the relative part is
Taub-NUT space.

%-------------------------------------------------------------------------------
\subsection{Second BPS equation}
%-------------------------------------------------------------------------------

The secondary BPS equation around the first BPS equation can be found
also explicitly by the method summarized in the appendix of
Ref.~\cite{leeyi}.  Here we briefly outline the solution following the
reference.  The Nahm data for $SU(2)$ one caloron case is given
by~\cite{lu,kraan}
\beq
{\bf T}_0 = -{\bf a}_0\; ,\quad {\bf T}_1 = -{\bf a}_1\; ,
\eeq
where ${\bf a}_{0,1}$ are constant vectors representing the
positions of constituent monopoles. We can put ${\bf a}_{0,1}$ by
spatial rotation and translation at the $z$-axis such that ${\bf
a}^i_0 = -\frac{R}{2}\delta^{i3},{\bf
a}^i_1=\frac{R}{2}\delta^{i3}$. In ADHMN formalism the covariant
Laplacian for adjoint scalar field $\phi$ is replaced by the
equation for function $p(t)$
\beq
\ddot{p}(t)-W(t)p(t)+\Lambda(t)=0\; ,         \label{peq}
\eeq
where $W(t)$ and $\Lambda(t)$ are given by
\beq W(t)\equiv{\rm tr}_2   \sum_{a=1}^2 \delta(t-h_a)w^\dagger_a
w_a\; , \quad \Lambda = {\rm tr}_2  \sum_{a=1}^2 \delta(t-h_a)
w^\dagger_a \langle\phi\rangle w_a \; , \eeq
For the given  Nahm data, $w_{1,2}$ are
\beq
w_1= ( \sqrt{2R}, 0)\; , \quad w_2= ( 0, \sqrt{2R})\; ,
\eeq
which is determined by the jumping condition
\beq
{\bf T}(h_a+) - {\bf T}(h_a-) = \frac{1}{2}\tr_2(\sigma^iw^\dagger_aw_a) \; .
\eeq
The solution of Eq. (\ref{peq}) for these data and the boundary value of the
adjoint scalar field $\phi$
\beq \langle \phi \rangle = -\frac{\eta}{2}\sigma^3= {\bf a}
\cdot{\bf H}\; ,\qquad {\bf a}\equiv\eta\bbeta_1 \eeq
is given by
\beqn
p(t)&=&\left\{ \begin{array}{ll}
        p_0(t+{\pi \over \beta}),   & \qquad
             t\in [-{\pi\over\beta},-{v\over 2}]     \\
          p_1t,         &   \qquad  t\in [-{v\over 2},{v\over 2}]  \\
      p_0(t-{\pi \over \beta}), & \qquad t\in [{v\over 2},{\pi\over\beta}]
              \end{array}
               \right.
                     \; , \\
  p_0&=&-\eta\frac{{\beta\over 2\pi} v R}{1+(1-{\beta\over 2\pi}v)v R}
     \; ,  \\
 p_1&=&\eta\frac{(1-{\beta\over 2\pi}v)R}{1+(1-{\beta\over 2\pi}v)v R}\; .
\eeqn
Then, we can construct the adjoint scalar field
by
\beq \phi = N^{-1/2} \langle\phi\rangle N^{-1/2} + N^{-1/2} \int
dt u^\dagger p(t) u N^{-1/2} \; , \eeq
where $N$ and $u(t)$ are the functions given in the reference
~\cite{lu}. It is straightforward to obtain the $\phi$ field  in
terms of the functions given in the reference~\cite{lu} and
extract the asymptotic behaviour of $\phi$.

{}From the asymptotic behaviour, we get
\beq
\phi = -\eta \frac{\sigma_3}{2}\left( 1 -\frac{R}{r}
\frac{\beta}{\beta +2\pi\mu_0\mu_1 R} \right)+ {\cal O}(\frac{1}{r^2})
           \;  ,
\eeq
and thus the electric charge  is given by
\beq Q^E = \frac{4\pi}{e^2}  \frac{ \eta^2\beta^2
R}{\beta+2\pi\mu_0\mu_1 R}\; . \eeq
%

%-----------------------------------------------------------
\subsection{Moduli space metric and Tong's method}
%-----------------------------------------------------------

The moduli space of $SU(2)$ one caloron is 8 dimensional and this
 8 dimensional moduli space can be split into the center of motion
space and the relative motion space and their metrics are given
by~\cite{piljin,kraan,lu,kraan2}
\beqn
ds^2_{cm}& =& 8\pi^2 \left\{ (d{\bf X})^2 + (d\chi_{cm})^2 \right\}\; ,\\
ds^2_{\rm rel} &=& 8\pi^2 \mu_0\mu_1 \left\{ (1+ \frac{\bar{r}}{r} ) d{\bf r}^2
+ \bar{r}^2 (1+ \frac{\bar{r}}{r})^{-1} (d\psi +
{\bf w}({\bf r}) \cdot d {\bf r} )^2 \right\} \; ,
\eeqn
%
% The metric of the 4-dimensional
%moduli metric for the center of the mass motion is flat and is given
%by
%
%\beq
%ds_{c.m.}  = d {\bf X}^2 + d \chi_T^2
%\eeq
%
%The relative motion metric is
%given by
%
%\beq
%ds^2_{\rm rel} = 8\pi^2 \mu_0\mu_1 \left\{ (1+ \frac{\bar{r}}{r} ) d{\bf r}^2
%+ \bar{r}^2 (1+ \frac{\bar{r}}{r})^{-1} (d\chi +
%{\bf w}({\bf r}) \cdot d {\bf r} )^2 \right\}
%\eeq
%
where $\bar{r} = \frac{\beta}{2\pi \mu_0 \mu_1}$. The relative metric $ds_{rel}^2$ is the one for the
well-known Taub-NUT space.
For this case
\beq \frac{\beta}{e^2}{\bf q} =\frac{1}{e^2} g_{\psi\psi}({\bf a}
\cdot{\bf K}^\psi) {\bf K}^\psi \; . \eeq
Since $q_\psi = q_1-q_0, {\bf q}=q_\psi\bbeta_1$ and ${\bf K}^\psi=\bbeta_1$,
one can see that
\beqn \frac{\beta}{e^2} q_\psi & =& \frac{1}{e^2}g_{\psi\psi}({\bf
a} \cdot {\bf K}^\psi)
=\frac{\eta}{e^2} g_{\psi\psi} \\
&=& \frac{4\pi}{e^2} \frac{\eta\beta^2 R}{\beta + 2\pi\mu_0\mu_1
R} \; . \eeqn
This gives us the electric charge $Q^E$
\beq Q^E = \frac{\beta}{e^2} {\bf a}\cdot {\bf q}=
\frac{4\pi}{e^2}  \frac{\eta^2\beta^2 R}{\beta+2\pi\mu_0\mu_1 R}
\; ,
\label{esu2}
\eeq
which is the identical expression with one obtained in the previous section.

The low energy dynamics of this configuration is described  by the
free c.m. part and the relative Taub-NUT metric with potential
term $U = \frac{1}{2}Q^E$. Note that the field theory BPS energy
bound~(\ref{bpsbound}) is saturated by ${\cal T}=
\frac{8\pi^2}{e^2}$ and the above $Q^E$.

%-----------------------------------------------------------------------
\subsection{String Picture}
%-----------------------------------------------------------------------

A dyon in maximally supersymmetric four-dimensional  Yang-Mills
theory appears as fundamental and D-string composition connecting
D3-branes. The positions of D3-branes are specified by the adjoint
scalar vacuum expectation values and the total energy of strings
connecting D3-branes matches with the field theoretic energy of
dyons~\cite{bergman}. In caloron case the string picture appears
after the T-dual is taken and then the vacuum expectation value of
$A_4$ specify the position of D3-branes.

With the tension for a single fundamental string $T=1/2\pi\alpha'$,
the tension for the D-string is $gT$ with $g=4\pi\beta/e^2$ in terms
of the field
theory parameters. The tension of
  $s = \beta q/e^2$ fundamental strings on a single
D-string is $\sqrt{g^2+s^2}T$, where $q$ is the field theory
electric charge.

A single dyonic caloron in the $SU(2)$ gauge group can be represented
in a the string picture by the T-dual transformation. Both $A_4$ and
$\phi$ expectation values denote the position of the D3 branes in the
compact dual space and transverse $x^5$ direction given by the $\phi$
field. The detail of the string picture of this configuration is given
below.

\vskip 5mm

\begin{center}
\leavevmode
\epsfysize=2in\epsfbox{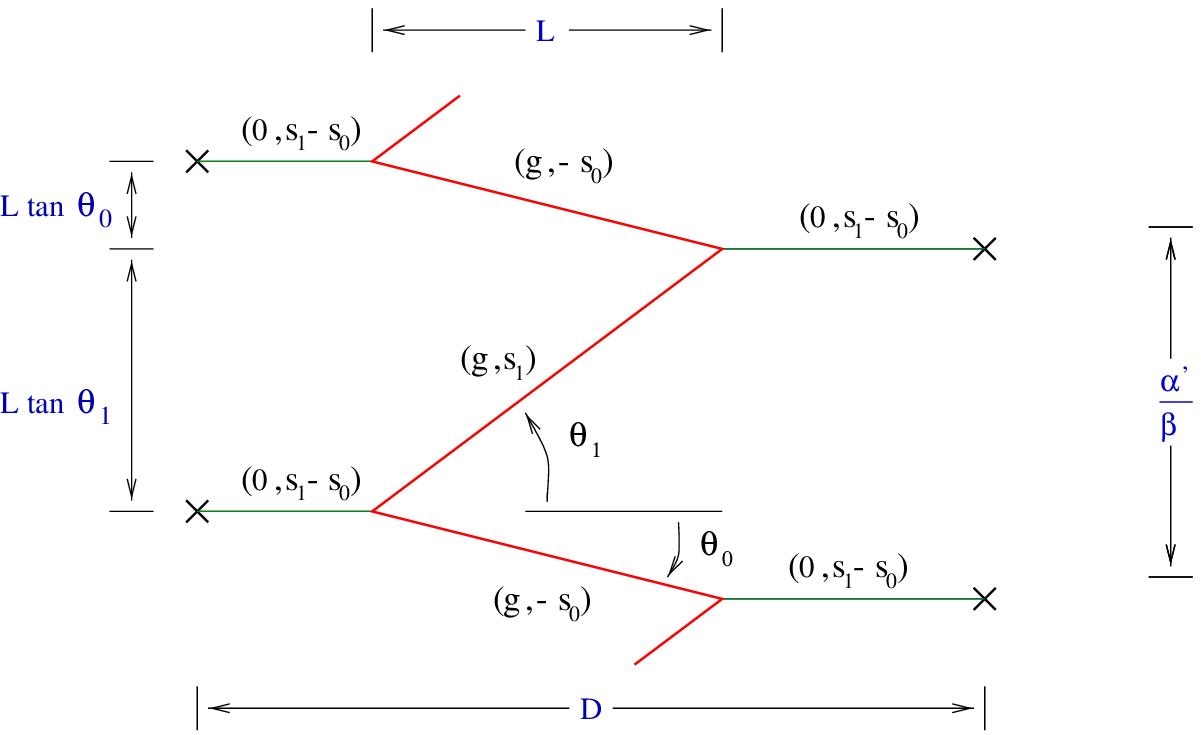}
\end{center}

\begin{quote}
{\bf Figure 1:} {\small String web picture of SU(2) single caloron
case}
\end{quote}

The above string configuration is 1/4 BPS when there is a tension
balance at the string junctions~\cite{sen}, which implies that
\beqn
&& \sin\theta_1 = \frac{g}{\sqrt{g^2+s_1^2}} \\
&& \sin\theta_0 = \frac{g}{\sqrt{g^2+s_0^2}} \eeqn
where $s_i=\frac{\beta}{e^2} q_i$.

Let us consider the total energy of the above string configuration,
which is the sum of the energy, that is the tension times the length,
of the individual string segments. The total string energy is then
\beq E =  (s_1-s_0) T  (D-L) + \sqrt{g^2 +s_1^2} T
\frac{L}{\cos\theta_1} + \sqrt{g^2+s_0^2}T \frac{L}{\cos\theta_0}
\; . \eeq
This can be written as
\beq E= g T L(\tan\theta_1+\tan\theta_0) + (s_1-s_0)T D \eeq
which is what we expect from the BPS energy from field theory
since we can identify the string theory parameters with the field
theory ones as $ TD =\eta$, $\beta T L \tan\theta_i=2\pi \mu_i,\
(i=0,1)$. The critical charge appears when $L=D$ or,
\beq \Delta \tilde{q}_c\equiv (s_1-s_0)_c =\frac{\beta
T}{2\pi\mu_1\mu_0}
 g D \eeq
which is identical with
$\lim_{R\rightarrow\infty}(q_1-q_0)\frac{\beta}{e^2}$ from the
field theory result (\ref{esu2}).

%-----------------------------------------------------------------
\section{Conclusion}
%-----------------------------------------------------------------

In this paper we have investigated 1/4 BPS configuration on
$R^{3+1}\times S^1$ in ${\cal N}=4$ supersymmetric Yang-Mills
theory. The generic BPS bound and BPS equations are obtained when
$P^4, Q^E, Q^M$ and ${\cal T}$ are turned on. The BPS equations
are complicated than 1/4 BPS configuration on $R^{3+1}$. Nevertheless,
BPS equations can be simplified by suitable transformation,
resulting in two equations. Then, the static solution for 1/4 BPS
configuration can be obtained by solving first primary equation
which is much studied 1/2 BPS caloron equation on $R^3\times S^1$
and next solving the secondary equation on this 1/2 BPS caloron
background.  This is quite similar to the  approach for 1/4 BPS
dyonic state in the four dimensional field theory~\cite{leeyi2}.

New feature in the 1/4 BPS dyonic caloron case compared to 1/4 BPS
dyon is the appearance of topological charge ${\cal T}$ and four
momentum $P^4$ in BPS bound. In the theory of the $SU(2)$ gauge group,
there is no 1/4 BPS dyons in four dimensions but there are 1/4 BPS
caloron configurations in five dimensions.  In the infinite $\beta$
limit these dyonic calorons become the dyonic instantons which were
studied before~\cite{tong,townsend}.

We have also considered the low energy dynamics of this 1/4 BPS
configuration and it turns out that it is described by the same
type non-linear $\sigma$-model Lagrangian with 1/4 BPS dyons. But
the potential term is more complicated by the existence of $P^4$
and $Q^M$.

There are several directions needed for further study. First, it
would be interesting to understand the $\alpha\neq 0$ case from
the low energy dynamics view point. We have argued that it is
related to the center of mass motion of low energy Lagrangian.

Second, it would be interesting to study the quantum spectrum of
this configuration. After the quantization the translational
generator along $x^4$ direction takes discrete integer values.
However the momentum $P^4$ could take fractional value. Thus, it
would be interesting to elucidate this point.

\newpage
\centerline{\bf Acknowledgments}

We thank Choonkyu Lee for useful discussions.  This work is supported
in part by the KOSEF 1998 Interdisciplinary Research Grant
98-07-02-06-01-5 (K.L.) and the BK21 program of Ministry of Education
and the Korea Research Foundation Grant 2001-015-DP--85 (S.-H. Yi).

\end{document}